\documentclass[american]{article}
\usepackage[T1]{fontenc}
\usepackage[latin9]{inputenc}
\usepackage{geometry}
\usepackage{amsmath}
\usepackage{amsfonts,latexsym,eucal,amssymb,dsfont}
\usepackage{mathtools}
\usepackage{graphicx}
\usepackage[authoryear]{natbib}
\usepackage{float}
\usepackage{url}
\usepackage{authblk}
\usepackage{tikz}
\usepackage{libertine}
\usepackage{subfig}

\def\dd{\textrm{d}}
\usepackage[final]{changes}

\usepackage{fullpage}
\begin{document}


\title{Modeling the impact of birth control policies on China's
  population and age: effects of delayed births and minimum birth age
  constraints}

\author[$1^*$]{Yue Wang} \author[$2^*$]{Renaud
  Dessalles} \author[$1,3$]{Tom Chou}

\affil[$1$]{Dept. of Computational Medicine, UCLA, Los Angeles, CA
  90095-1766, USA}
\affil[$2$]{Greenshield, 46 rue Saint Antoine, 75004 Paris, 
France}
\affil[$3$]{Dept. of Mathematics, UCLA, Los Angeles, CA
  90095-1555, USA}
\affil[$^{*}$]{These authors contributed equally to this work}
\maketitle 

\begin{abstract}
We consider age-structured models with an imposed refractory period
between births. These models can be used to formulate alternative
population control strategies to China's one-child policy. By allowing
any number of births, but with an imposed delay between births, we
show how the total population can be decreased and how a relatively
older age distribution can be generated. This delay represents a more
``continuous'' form of population management for which the strict one-child
policy is a limiting case.  Such a policy approach could be more
easily accepted by society. Our analyses provide an initial framework
for studying demographics and how social constraints influence
population structure.
\end{abstract}


\vspace{3mm}

%
%
%
%
%
%
%

\section{Introduction}

Models of age-structured population dynamics are often based on the
classic McKendrick equation \citep{MCKENDRICK,Kermack1927} (sometimes
called von Foerster equation \citep{VONFOERSTER}).  These equations
describe the dynamics of the mean population as a function of time $t$
and expressed as a density in age $a$. The solutions to the McKendrick
equations can be partially solved using the method of characteristics
and numerical approximations \citep{Perthame2007,KEYFITZ} across many
contexts. Moreover, stochastic extensions to incorporate the random
times of birth and death (demographic stochasticity) have been
formulated using branching processes \citep{jiang2017phenotypic} and
kinetic and operator theory
\citep{Greenman2016,Greenman2016_JSP,Greenman2017,kinetic_theory}

Age-structured equations have been used to predict the evolution of
human and animal populations
\citep{tuljapurkar1983transient,bongaarts1985alternative,feeney1987period}.
Using such models and ideas from control theory to frame population
control strategies was vogue in the 1970s
\citep{LANGHAAR1973,POLLARD1973,FALKENBURG1973,HRITONENKO}.  A
profound example was its use in 1979 by Jian Song
\citep{JSONG1980,JSONG1982}, a Chinese engineer who numerically solved
the one-component McKendrick equation using birth rates associated
with China in the late 1970s (see
Fig.~\ref{fig:example_birth-death-rates}). By projecting future
populations associated with different birth rates (expressed by the
mean number of children per woman), he found that in order to keep the
population manageable ($\sim$ 700 million - 1 billion) within 100
years, this control parameter would have to be decreased to the point
where each woman is allowed only one child
\citep{JSONG1980,JSONG1982,JSONG1988}. This research provided the
technical basis of the one-child policy in China
\citep{Greenhalgh2004,Bacaer2011}.


In the 1970s, China had encouraged (but not enforced) people to marry
later, wait longer before childbearing, and have fewer children
(``later-longer-fewer'' policy) \citep{bongaarts1985alternative}.
Despite concerns from social scientists and demographers who proposed
such ``softer'' controls, the one-child policy was implemented in
1980, based on the implications of Jian Song's numerical solutions to
the McKendrick equation. Rather than imposing a maximal number of
children, a minimum delay between two consecutive births
\citep{Greenhalgh2004} or a minimum birth age could have been imposed.
Such a policy would arguably have been more easily enforced and would
have led to fewer unintended consequences such as a skewed sex ratio
and an elder-heavy age distribution. Here, we retrospectively model
such alternatives and make predictions as policies change.


Specifically, we extend the McKendrick age-structured model to
incorporate a delay between successive births by each
female. \added{In order to do so, we must explicitly delineate
  individuals who have not given birth from those who have given birth
  at least once. Imposed delays between successive births can then be
  formally described by adjusting the birth rate function of the
  individuals who have given birth at least once. We solve our model}
equations using parameters appropriate to 1981 China and compare
predictions of the graded policies with those of a strict one-child
policy. We explore how the total population and age distribution are
affected by different values of imposed refractory periods and minimum
birth age.

\section{Mathematical Model}
\label{sec2}
When applying age-structured partial differential equation (PDE)
models to two-sex populations, a simple assumption is to consider only
the density of females at time $t$ with age $a$.  The
\added{predicted} number of females with age between $a$ and $a+\dd a$
is thus $f(a,t)\dd a$.  Indeed, unless the female population is much
larger than the male population (e.g., after a war), the female
population can be considered as the ``limiting quantity'' that
determines the number of births.  In other words, the frequency of
births in the total population is relatively insensitive to the male
population. The McKendrick equations describing the female population
density $f(a,t)$ are formulated as

\begin{subequations}
\label{eq:Sys0} 
\begin{align}
\frac{\partial}{\partial t}f(t,a)+
\frac{\partial}{\partial a}f(t,a)=-\mu_{\rm f}(a)f(t,a) \label{eq:Sys0_evol}\\
f(t,0)=\eta \int_{0}^{\infty}\!\!\beta_{\text{eff}}(a) f(t,a)\dd a
\label{eq:Sys0_limCond}\\
f(0,a)=I_{\rm f}(a),\label{eq:eq:Sys0_init}
\end{align}
\end{subequations}
where $\mu_{\rm f}(a)$ represents the death rate of females of age
$a$, $\beta_{\text{eff}}(a)$ is the observed birth rate of women of
age $a$, $\eta$ is the fraction of births that produce girls,
\added{and $I_{\rm f}(a)$ is the age distribution of the initial
  population at $t=0$}.

Equation~(\ref{eq:Sys0_evol}) describes the time-evolution of the
population, Eq.~(\ref{eq:Sys0_limCond}) denotes the boundary condition
at age $a=0$ describing the number of girls born at time $t$, and
Eq.~(\ref{eq:eq:Sys0_init}) specifies the initial condition. This
model neglects the explicit mating-age male population which is valid
when $\eta$ is maintained below $0.5$, giving rise to more males than
females. \added{For humans, $\eta\approx 0.48-0.49$ naturally (but this is
  compensated by a slightly higher mortality in males across all
  ages).} With sex-selective abortion, $\eta$ can be even smaller
\citep{RATIO2014}.  If one were also interested in the male population
$m(t,a)$, it would obey the same equations except with a male version
of the death rate $\mu_{\rm m}(a)$, an initial condition $I_{\rm
  m}(a)$, and a boundary condition for male newborns:
$m(t,0)=(1-\eta)\int_{0}^{\infty}\beta_{\rm eff}(a)f(t,a)\dd a$.


\subsection{Delayed birth model}

Now, in order to introduce a delay between consecutive births, we need
to further partition the female population into those who have never
had a child and those who have already had a child (and who may need
to wait a certain time before having another one). The population
densities for each of these classes of females are defined as:
\begin{description}
\item [{$f_{0}(t,a)$:}] the population density of childless
  females. The quantity $f_{0}(t,a)\dd a$ is the number of females with
  age between $a$ and $a+\dd a$ and who have never had a child up to
  the current time $t$.

\item [{$f(t,a,\tau)$:}] the population of females who have had at least
  one child. The quantity $f(t,a,\tau)\dd a \dd\tau$ is the number of
  females at time $t$ whose age is between $a$ and $a+\dd a$ and whose
  youngest child's age is between $\tau$ and $\tau+\dd\tau$.
\end{description}
We will assume that these two populations have the same age-dependent
death rate $\mu_{\rm f}(a)$ but give birth at different rates
$\beta_{0}(a)$ and $\beta(a,\tau)$, respectively.  We also define the
total female population density as

\begin{equation}
f_{\rm tot}(t,a)=f_{0}(t,a)+\int_{0}^{a}\!f(t,a,\tau)\dd \tau
\label{eq:def_f_tot}
\end{equation}
and the total number of females at time $t$ as
\begin{equation}
n(t)=\int_{0}^{\infty}\!\!f_{\rm tot}(t,a)\dd a
=\int_{0}^{\infty}\!\!f_{0}(t,a)\dd a+
\int_{0}^{\infty}\!\! \dd a \int_{0}^{a}\!\!\dd\tau  f(t,a,\tau).
\label{NTOT}
\end{equation}

The age-structured McKendrick equations for $f_{0}$ and $f$ are:
\begin{subequations}
\label{eq:Sys1} 
\begin{align}
\frac{\partial}{\partial t}f_{0}(t,a)+\frac{\partial}{\partial a}f_{0}(t,a)
& = {}-\left(\mu_{\rm f}(a)+\beta_{0}(a)\right)f_{0}(t,a)\label{eq:evol0} \\
\frac{\partial}{\partial t}f(t,a,\tau)+\frac{\partial}{\partial a}f(t,a,\tau)
+\frac{\partial}{\partial\tau}f(t,a,\tau) & = {}-\left(\mu_{\rm f}(a)
+\beta(a,\tau)\right)f(t,a,\tau)\label{eq:evol1}
\end{align}
\begin{align}
f_{0}(t,0) & =\eta\left(\int_{0}^{\infty}\!\!\beta_{0}(a)f_{0}(t,a)\dd a
+ \int_{0}^{\infty}\!\!\dd a \int_{0}^{a}\!\!\dd\tau\, \beta(a,\tau)f(t,a,\tau)  \right)  
\label{eq:limCond0}\\
 f(t,a,0) & =\beta_{0}(a)f_{0}(t,a)+ \int_{0}^{a}\!\!\beta(a,\tau)f(t,a,\tau)\dd\tau  
\label{eq:limCond1}\\
 f_{0}(0,a) & =I_{0}(a) \quad\text{and}\quad f(0,a,\tau)=I(a,\tau). \label{eq:init}
\end{align}
\end{subequations}
Equation~(\ref{eq:evol0}) describes the evolution of $f_{0}$ as in the
classical McKendrick equation (cf Eq.~(\ref{eq:Sys0_evol})) with birth
rate $\beta_{0}(a)$.  For $f(t,a,\tau)$, we must introduce the new
variable $\tau$ to mark the time since the last birth.  This brings in
another convection term in Eq.~(\ref{eq:evol1}) since $\tau$ increases
alongside time $t$ and age $a$. The birth rate $\beta$ for this
population can depend on both the age $a$ and the time $\tau$ since
the last birth.
Eq.~(\ref{eq:limCond0}) gives the number of girls $f_{0}(t,0)$ born at
time $t$, while Eq.~(\ref{eq:limCond1}) describes $f(t,a,0)$, the
density of females at age $a$ at time $t$ who just gave birth.
%
%
These individuals can arise from the $f_{0}$ population (females who
have never had a child) or from the $f$ population itself (females who
have already had at least one child). Thus, the boundary conditions
~(\ref{eq:limCond0}) and ~(\ref{eq:limCond1}) couple the two
populations $f_{0}$ and $f$.  Finally, equations~(\ref{eq:init})
simply describe the initial conditions for $f_{0}$ and $f$.
   
In Appendix \ref{ap:ftot}, we explicitly show that the total female
population density $f_{\rm tot}(t,a)$ (Eq.~\eqref{eq:def_f_tot})
satisfies the standard age-structured McKendrick equation

\begin{equation} 
\label{eq:f_tot}
	\begin{split}	 
\frac{\partial}{\partial t}f_{\rm tot}(t,a) 
+\frac{\partial}{\partial a}f_{\rm tot}(t,a)
=-\mu_{\rm f}(a)f_{\rm tot}(t,a).
\end{split}
\end{equation}
\added{Within a model that explicitly considers the time $\tau$ since
  the last childbirth, we can easily describe an imposed hypothetical
  policy that applies a refractory period $\delta$ between
  births. After having a child and before this refractory period
  ($0\le \tau\le \delta$) expires, the birth rate $\beta(a,\tau)$ can
  be set to $0$. As a preliminary description, we will consider a
  policy-modified (truncated) birth rate function}
\begin{equation}
\beta(a,\tau)=\beta_{0}(a)\mathds{1}(\tau,\delta)
\label{eq:beta}
\end{equation}
\added{where} the indicator function $\mathds{1}(\tau,\delta) = 1$ for
$\tau > \delta$ and $\mathds{1}(\tau,\delta) = 0$ for $\tau \leq
\delta$.  This form assumes that once the imposed refractory period
has past, the birth rate immediately rises back to a value associated
with the persons current age.
   
%

\subsection{Asymptotic behavior}

We first analyze the asymptotic behavior of our model. An important
feature of renewal transport equations such as the McKendrick model is
that as $t\to \infty$, the total population $n(t)$ will grow
exponentially (in the absence of nonlinear regulation terms
\citep{MACCAMY1974}), while the normalized, age-dependent population
density converges to a time-independent stationary distribution (see
\citet[Chapter 3]{Perthame2007} and \citet{Arino1995}). This property
is independent of the initial condition.  We will assume that this
steady-state asymptotic property arises in our two-component,
three-variable model; \textit{i.e.}, the normalized densities
$f_{0}(t,a)/n(t)$ and $f(t,a,\tau)/n(t)$ converge to stationary
distributions.  We denote the stationary limits as
\begin{equation}
\lim_{t\to \infty} \frac{f_{0}(t,a)}{n(t)} \equiv h_{0}(a)
\quad \text{and}\quad \lim_{t\to \infty}\frac{f(t,a,\tau)}{n(t)} \equiv h(a,\tau).
\label{eq:def_h}
\end{equation}
We also define the distribution associated with the total 
female population as
\begin{equation}
 \lim_{t\to \infty}\frac{f_{\rm tot}(t,a)}{n(t)}
\equiv h_{\rm tot}(a) = h_{0}(a)+\int_{0}^{a}\! h(a,\tau)\dd\tau,
\label{eq:def_h_tot}
\end{equation}
where $\int_{0}^{\infty} h_{\rm tot}(a) \dd a=1$.  If we assume that
$f_{0}(0,a)/n(0)=h_{0}(a)$ and $f(0,a,\tau)/n(0)=h(a,\tau)$ for any
$a,\tau$ at some initial time $t=0$, then
$f_{0}(t,a)/n(t)=h_{0}(a)$ and $f(t,a,\tau)/n(t)=h(a,\tau)$ hold for
any $t\ge 0$.

From Eq.~(\ref{eq:evol0}), we have
\begin{equation}
\begin{aligned} 
\label{eq:lambda}			 
\frac{1}{f_{0}(t,a)}
\frac{\partial f_{0}(t,a)}{\partial t} & =  
-\frac{1}{f_{0}(t,a)}
\frac{\partial f_{0}(t,a)}{\partial a}-\mu_{\rm f}(a)  \\
\: & =  -\frac{n(t)}{f_{0}(t,a)}\frac{1}{n(t)}
\frac{\partial f_{0}(t,a)}{\partial a}
-\mu_{\rm f}(a) \\
\: & = \frac{1}{h_0(a)}\frac{\dd h_0(a)}{\dd a}-\mu_{\rm f}(a).
\end{aligned}			
\end{equation}
Thus, $[\partial f_0(t,a)/\partial t]/f_0(t,a)$ is independent of $t$.
Moreover, for any $a,a',\tau$
\begin{equation} 
\label{eq:ratio}			 
\frac{f_0(t,a)}{f(t,a',\tau)}
=\frac{h_0(a)}{n(t)}\frac{n(t)}{h(a',\tau)}
=\frac{h_0(a)}{h(a',\tau)}			
\end{equation}
is also independent of $t$ so that
\begin{equation} 
\label{eq:lambda2}			 
\frac{\partial}{\partial t}\left[\frac{f_0(t,a)}{f(t,a',\tau)}\right]=
\frac{1}{f(t,a',\tau)^2}\left[f(t,a',\tau)\frac{\partial}{\partial t}f_0(t,a)
-f_0(t,a)\frac{\partial}{\partial t}f(t,a',\tau)\right]=0.			
\end{equation}
Thus, for any $a,a',\tau$,
\begin{equation} \label{eq:lambda3}
    \frac{1}{f_{0}(t,a)}\frac{\partial f_{0}(t,a)}{\partial t} =
    \frac{1}{f(t,a',\tau)}\frac{\partial f(t,a',\tau)}{\partial t}
=\frac{1}{f_{0}(t,a')}\frac{\partial
      f_{0}(t,a')}{\partial t}.
%
\end{equation}
Eqs.~(\ref{eq:lambda}) and (\ref{eq:lambda3}) show that $[\partial
  f_0(t,a)/\partial t]/f_0(t,a)$ is independent of both $t$ and $a$,
allowing us to define a constant that describes the stationary growth
rate
\begin{equation} 
\label{eq:lambda4}			 
\lambda = \frac{1}{f_{0}(t,a)}\frac{\partial f_{0}(t,a)}{\partial t} =
\frac{1}{f(t,a,\tau)}\frac{\partial f(t,a,\tau)}{\partial t} =
\frac{1}{f_{\rm tot}(t,a)}
\frac{\partial f_{\rm tot}(t,a)}{\partial t}.		
\end{equation}
%
%
%
Thus, we can express solutions for the densities $f_{0}(t,a)$ and
$f(t,a,\tau)$ in the form
\begin{equation}
f_{0}(t,a)= C h_{0}(a)e^{\lambda t}\quad\text{and}
\quad f(t,a,\tau)= C h(a,\tau)e^{\lambda t},
\label{eq:fh_lambda}
\end{equation}
where $C$ is a constant. After using these expressions in
Eqs.~(\ref{eq:Sys1}), we find the equations for the stationary
distributions
\begin{subequations}
\label{eq:Sys2}
\begin{align}
\frac{\dd}{\dd a}h_{0}(a) & =-\left(\mu_{\rm f}(a)+\beta_{0}(a)+\lambda\right)
h_{0}(a)
\label{eq:evol0-1}\\
\frac{\partial}{\partial a}h(a,\tau)+\frac{\partial}{\partial\tau}h(a,\tau) & 
=-\left(\mu_{\rm f}(a)+\beta(a,\tau)+\lambda\right)h(a,\tau)
\label{eq:evol1-1}\\
h_{0}(0) & =\eta \left(\int_{0}^{\infty}\beta_{0}(a)h_{0}(a)\dd a+
\int_{0}^{\infty}\!\! \dd a \int_{0}^{a}\!\dd \tau\, \beta(a,\tau)h(a,\tau)\right)
\label{eq:limCond0-1}\\
h(a,0)& =\beta_{0}(a)h_{0}(a)+\int_{0}^{a}\!\beta(a,\tau)h(a,\tau)\dd \tau.
\label{eq:limCond1-1}
\end{align}
\end{subequations}
Next, using Eq.~(\ref{eq:f_tot}), we find
\begin{equation}
\frac{\dd}{\dd a}h_{\rm tot}(a) =-\left(\mu_{\rm f}(a)
+\lambda\right) h_{\rm tot}(a),
\label{eq:h_lambda0}
\end{equation}
which is solved by
\begin{equation}
h_{\rm tot}(a) =h_{\rm tot}(0)\exp
\left[-a\lambda-\int_0^a\!\! \mu_{\rm f}(a')\dd a'\right].
\label{eq:h_lambda}
\end{equation}
We can then define the effective whole-population birth rate function
\begin{equation}
\beta_{\rm eff}(a)=\frac{\beta_{0}(a)h_{0}(a)+\int_{0}^{a}\!
\beta(a,\tau)h(a,\tau)\dd\tau}{h_{\rm tot}(a)},
\label{eq:b_eff}
\end{equation}
which describes the overall birthrate weighted over the entire
stationary population. This population-averaged birth rate $\beta_{\rm
  eff}(a)$ corresponds to that used in the basic lumped model
(Eq.~\eqref{eq:Sys0_limCond}) and is the quantity that can be directly
extracted from birth data that provide women's ages at time of birth,
but that may not distinguish whether females are first-time mothers.
We prove in Appendix \ref{ap:beb0} that given
  $\beta_{\text{eff}}(a)$, the new-mother birth rate function can be
  calculated from
\begin{equation}
\beta_{0}(a)=\frac{\beta_{\text{eff}}(a)}{1-\int_0^\delta
\beta_{\text{eff}}(a-\tau)\dd \tau},
\label{eq:b0be}
\end{equation}
which then allows us to reconstruct $\beta(a,\tau)$ from
Eq.~(\ref{eq:beta}).  Using $\beta_{\rm eff}$, the boundary condition
for Eq.~\eqref{eq:h_lambda0}, the counterpart to
Eq.~(\ref{eq:limCond0-1}), can be written as
\begin{equation}
h_{\rm tot}(0)=\eta \int_0^{\infty} \beta_{\rm eff}(a)h_{\rm tot}(a)\dd a.
\label{eq:h0}
\end{equation}
Finally, after using the solution in Eq.~\ref{eq:h_lambda} for $h_{\rm
  tot}(a)$ in Eq.~(\ref{eq:h0}), we find an equation for $\lambda$:
\begin{equation} 
z(\lambda)\equiv \eta \int_0^{\infty} \beta_{\rm eff}(a)
    \exp\left[-a\lambda-\int_0^a \mu_{\rm f}(a')\dd a'\right]\dd a =1.
\label{eq:solvel}
\end{equation}
The function $z(\lambda)$ is monotonically decreasing with $\lambda$
and obeys the limits $\lim_{\lambda \to +\infty} z(\lambda) = 0$ and
$\lim_{\lambda \to -\infty} z(\lambda) = +\infty$. Thus,
Eq.~(\ref{eq:solvel}) has a unique solution that can easily be found
numerically.  From Eq.~(\ref{eq:solvel}), the solution for
$\lambda$--the net population growth rate--clearly increases with
$\beta_{\rm eff}(a)$ and decreases with $\mu_{\rm f}(a)$.

%
%

With $\beta_{0}(a)$ and $\lambda$ determined by Eqs.~ \eqref{eq:b0be}
and \eqref{eq:solvel}, respectively, we can explicitly find
$h(a,\tau)$.  First, we use the normalization condition $\int_0^\infty
h_{\rm tot}(a)\dd a=1$ on Eq.~\eqref{eq:h_lambda0} to explicitly find
$h_{\rm tot}(0)$ in terms of $\lambda$ and $\mu_{\rm f}(a)$.  Since
$h(0,\tau)=0$, we have $h_0(0)=h_{\rm tot}(0)$, which allows us to
explicitly express the solution to Eq.~(\ref{eq:evol0-1}):

\begin{equation} 
h_{0}(a) =h_{0}(0)\exp\left[-a\lambda
-\int_0^a \big(\mu_{\rm f}(a')+\beta_{0}(a')\big)\dd
      a'\right].
\label{eq:h0a}
\end{equation}
Next, we use Eq.~(\ref{eq:limCond1-1}) and Eq.~(\ref{eq:b_eff}) to
eliminate $\beta(a,\tau)$ and find $h(a,0)=h_{\rm
  tot}(a)\beta_{\text{eff}}(a)$, which is known.  Thus, we can
explicitly calculate $h(a,\tau)$ by solving Eq.~(\ref{eq:evol1-1})
using the method of characteristics:
\begin{equation} 
h(a,\tau) =h(a-\tau,0)\exp\left[-\tau\lambda-\int_{a-\tau}^a
      \big(\mu_{\rm f}(a')+\beta(a',a'-a+\tau)\big)\dd a'\right].
\label{eq:hat}
\end{equation}
To summarize, starting from $\beta_{\text{eff}}(a), \mu_{\rm
  f}(a),\eta$ (measured, say), and $\delta$, we can compute the growth
rate $\lambda$ numerically, then analytically reconstruct
$\beta_{0}(a),\beta(a,\tau), h_0(a)$ and $h(a,\tau)$.

\begin{figure}[htb]
	\centering{}\includegraphics[width=3.3in]{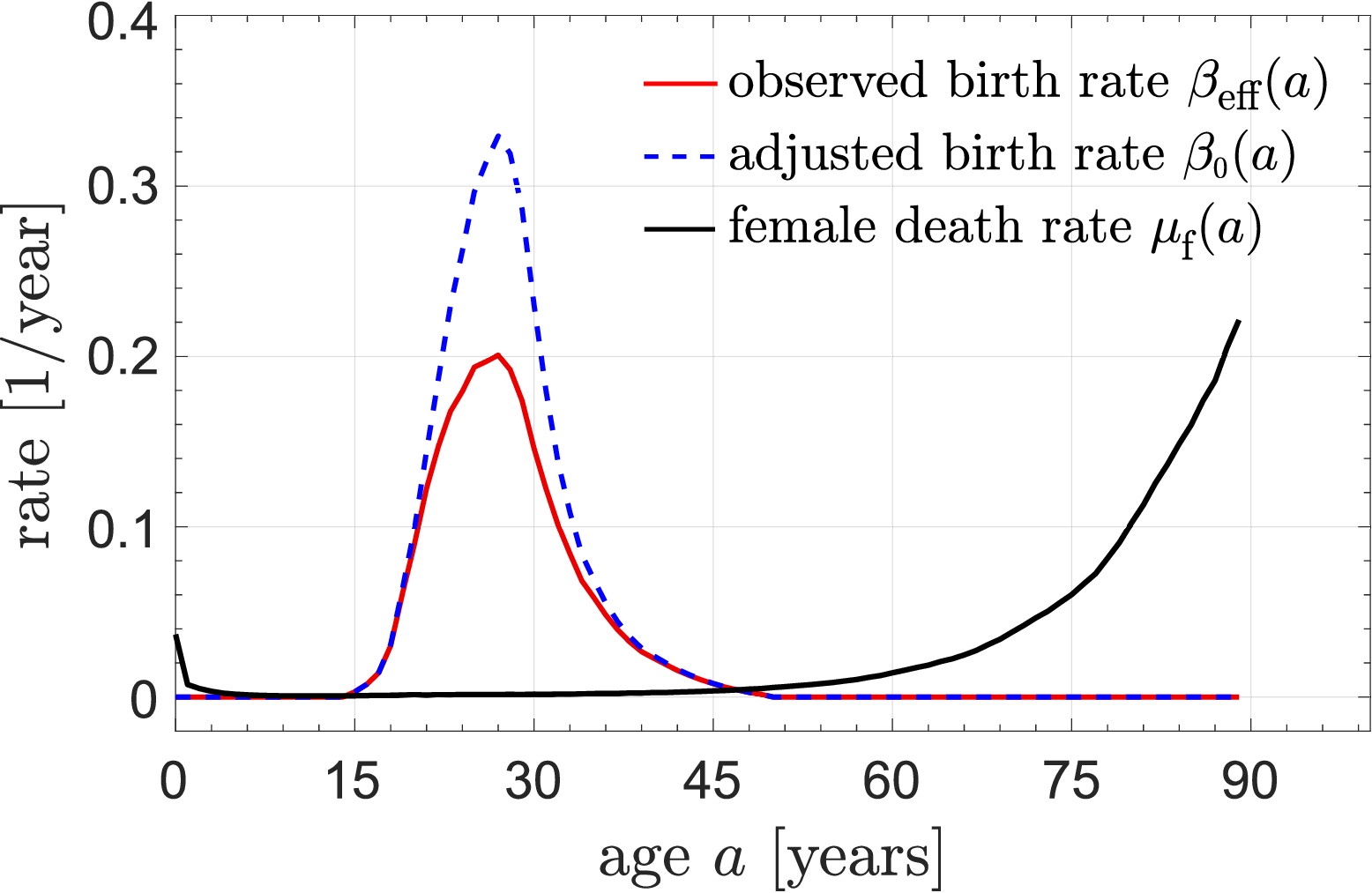}
	\caption{\label{fig:example_birth-death-rates} China's 1981
          birth rate and female death rate $\mu_{\rm f}(a)$,
          calculated from 1982 national census data
          \citep{Census1982}. The red curve represents the observed
          birth rate $\beta_{\rm eff}(a)$ for all women of a given
          age. The dashed blue curve represents the birth rate
          $\beta_{0}(a)$ for females who have not had any
          children. The area under the observed birth rate
          $\int_{0}^{\infty}\!\!\beta_{\rm eff}(a)\dd a$ represents
          the mean number of children born during over an individual's
          lifetime.}
\end{figure}

We now use the Chinese national census data recorded in 1982
\citep{Census1982} to infer the overall birth rate $\beta_{\rm
  eff}(a)$ and the female death rate $\mu_{\rm f}(a)$ functions in
China in 1981.  Although gestation imposes a hard refractory period of
$\delta \approx 9$ months, some time is needed to recover from
childbirth and the birth rate should more gradually recover.
\added{Specifically, in 1981 China, extended breastfeeding was common,
  which prevents the next pregnancy \citep{xu2009breastfeeding}. Thus,
  we will assume the birth rate returns to normal approximately only
  after about two years.  Therefore, when there is no policy that
  controls the interval between births, we set $\delta=2$ years such
  that $\beta(a,\tau)=0$ for $\tau < 2$ years and
  $\beta(a,\tau)=\beta_{0}(a)$ for $\tau>2$ years.  For other
  societies, this natural refractory period might be shorter. Using
  $\delta = 2$ and Eq.~(\ref{eq:b0be}), we calculate $\beta_{0}(a)$
  from $\beta_{\text{eff}}(a)$ derived from data. These rates are
  illustrated in Fig.~\ref{fig:example_birth-death-rates}.  Note that
  $\beta_0(a)$ for 1981 has already been mildly affected by the
  incipient birth-control policies in China.}

\begin{figure}[H]
\centering{}
\includegraphics[width=5.4in]{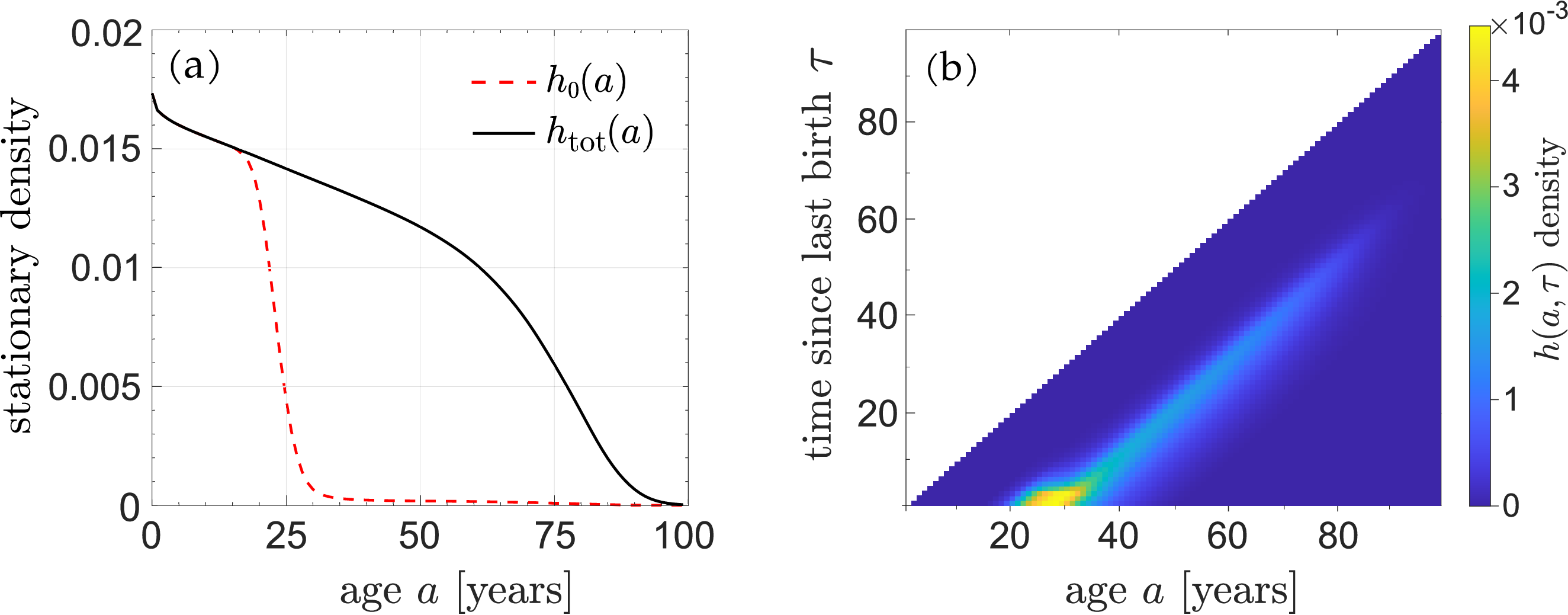}
\caption{\label{fig:example}The asymptotic population distribution
  associated with the birth and death rates shown in
  Fig.~\ref{fig:example_birth-death-rates}. (a) Steady-state age
  distributions of females without children, $h_{0}(a)$, and all
  females, $h_{\rm tot}(a)$, respectively. A monotonically decreasing
  $h_{\rm tot}(a)$ indicates that there are larger numbers of younger
  females, consistent with an increasing population and $\lambda > 0$.
  (b) The full double density $h(a,\tau)$ of females of age $a$ and
  whose youngest child is $\tau$ years old.}
\end{figure}

\begin{figure}[H]
\centering{}\includegraphics[width=5.4in]{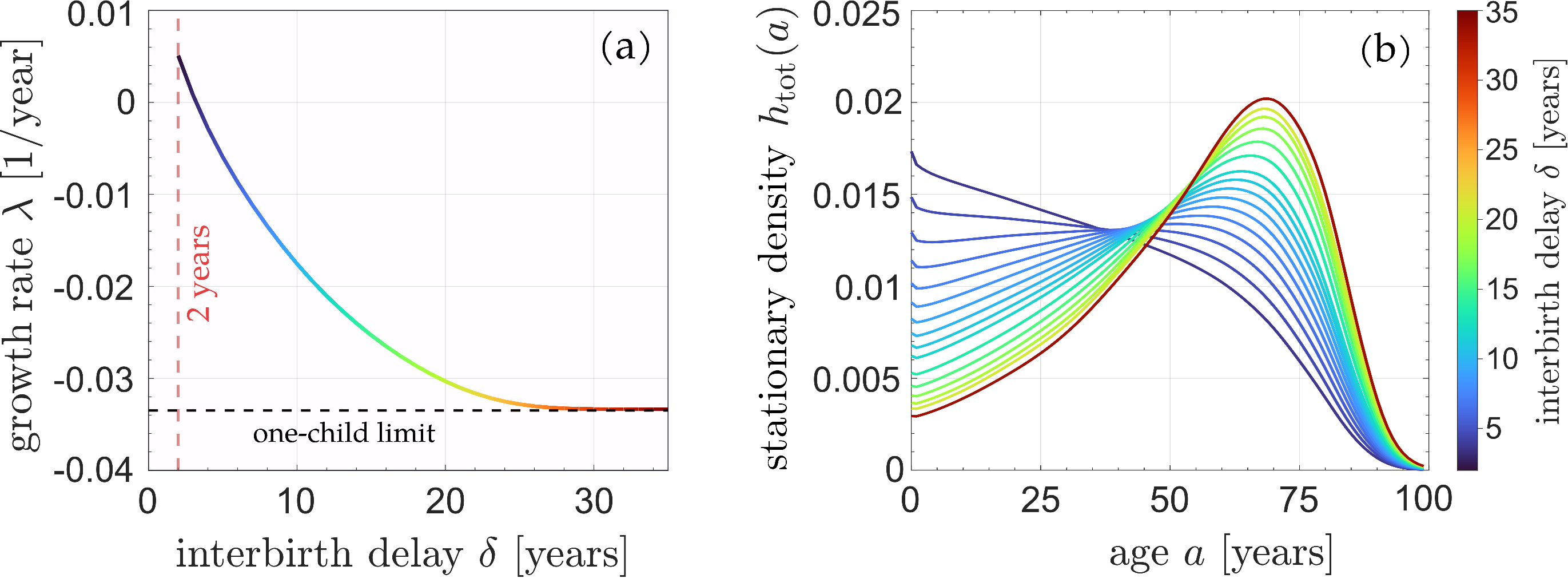}
\caption{\label{fig:delays} Asymptotic total-population growth rate
  and the steady-state, total-population age distribution.  (a) Effect
  of an imposed interbirth delay $\delta$ on the asymptotic growth
  rate $\lambda$. Most of the decrease in $\lambda$ occurs at small
  values of $\delta$ where the most negative slopes arise. When
  $\delta$ is large enough, a woman ages out before giving birth
  again, and imposing interbirth delay is equivalent to the strict
  one-child policy. (b) The steady-state age distribution $h_{\rm
    tot}(a)$. As $\delta$ increases, the peak of population density
  moves from $0$ to approximately $65$.}
\end{figure}

Using the birth rate $\beta_{\rm eff}(a)$ and death rate $\mu_{\rm
  f}(a)$ shown in Fig. ~\ref{fig:example_birth-death-rates}, we solve
Eqs.~(\ref{eq:Sys2}) to find $h_{0}(a)$ and $h(a,\tau)$, and plot them
with $h_{\rm tot}(a)$ given by Eq.~\eqref{eq:h_lambda} \added{in
  Fig.~\ref{fig:example}(a,b)}.
To explore the effects of an imposed refractory period, we first set
$\delta=2$ years, apply the newborn sex ratio of China in 1981,
$\eta=0.48$, and solve Eq.~\eqref{eq:solvel}. We find $\lambda \simeq
0.005 > 0$, indicating an exponentially growing total population.
This stationary growth rate is much smaller than the actual growth
rate of China in 1981, which is 0.0146. One reason is that in 1981,
the proportion of younger females is much higher than that in the
stationary distribution $h_{\rm tot}$.  The shape of $h_{\rm tot}(a)$
is consistent with this growth as it is monotonically decreasing,
indicating that every new generation has a larger population than the
previous one.  In Fig.~\ref{fig:delays}, we see how increases in the
refractory period $\delta$ decrease the asymptotic growth rate
$\lambda$ and affect the distribution $h_{\rm tot}(a)$. A negative
overall birth rate $\lambda<0$ (\textit{i.e.}, an asymptotically
decaying population) arises when $\delta \gtrsim 3.22$~years $\approx
39$~months.  As soon as $\lambda<0$, the distribution $h_{\rm tot}(a)$
becomes nonmonotonic, and a peak in the female population distribution
arises at a finite age $a>0$. 

\added{If $\delta$ is set sufficiently large, a female cannot have a
  second child, and the outcome is equivalent to a strict one-child
  policy. We use the terminology ``strict one-child policy'' for the
  scenario in which each female can have strictly no more than one
  child, while we use ``one-child policy'' to refer to the actual
  policy realized in practice. From 1980 to 1990, the one-child policy
  contained many exceptions, allowing one to bear more than one child
  \citep{nie1999problem}.}
%
%
Our formulation is valid only in the asymptotic case with a fixed
delay $\delta$ that remains unchanged for a long period of time.  For
practical modeling of policies in which delays $\delta$ are used as a
time-dependent control variable, such as China's 1980 one-child policy
and its subsequent modification in 2015, it is necessary to analyze
the full model that delineates the two female populations.

\subsection{Temporal evolution}

\begin{figure}
\centering{}\includegraphics[width=5.4in]{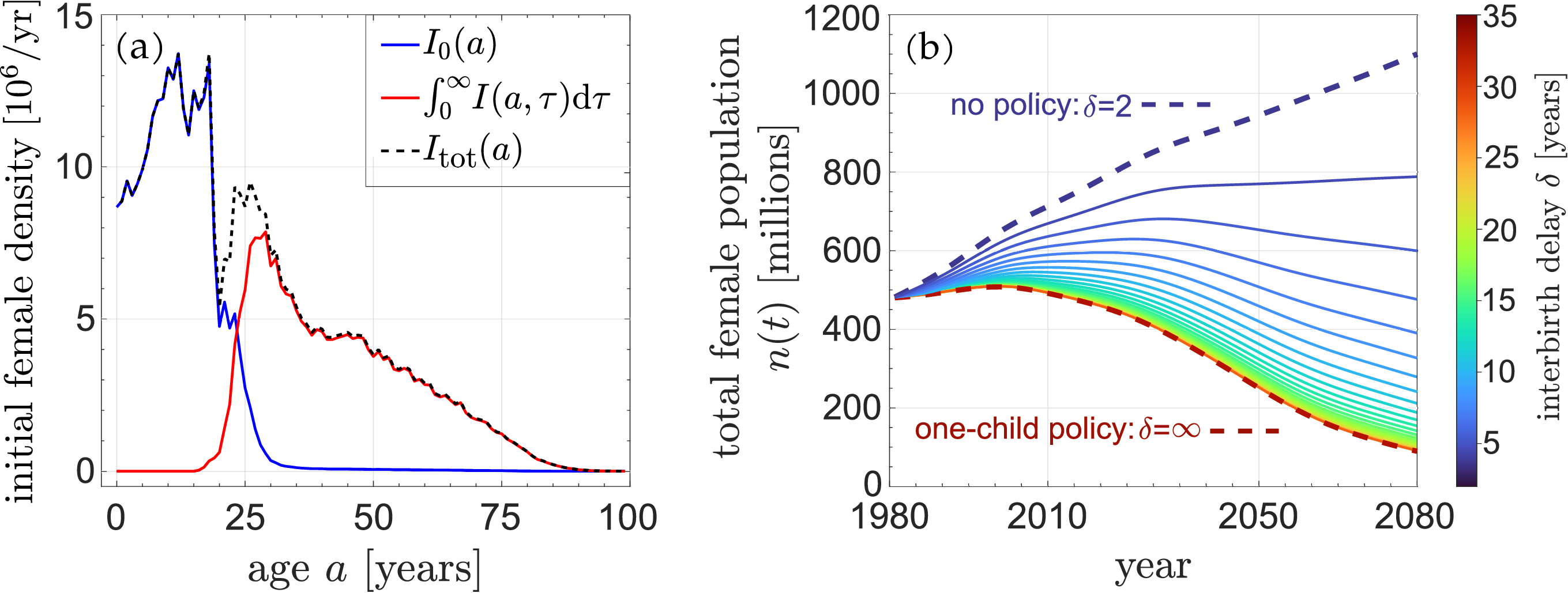}
\caption{\label{fig:several-delays}Evolution of the population for
  different delays. (a) Initial female subpopulation distributions in
  1981 China (calculated from \citet{Census1982}).  (b) Evolution of
  the total female population $n(t)$ for different imposed delays
  between births. These results are determined using the birth and
  death rates in Fig.~\ref{fig:example_birth-death-rates} and the
  initial populations shown in (a). A relatively small delay between
  births (for example, $\delta\sim 4$ years) has a significant impact
  on the evolution of $n(t)$.}
\end{figure}

As was used to predict the effects of the one-child policy, we use
China's female age distribution in 1981 \citep{Census1982} as a
starting point to explore how the total population evolves under
different values of the imposed delay $\delta$.  Since the data only
contain total female numbers $I_{\rm tot}(a)$ and not $I_{0}(a)$ and
$I(a,\tau)$ individually, we use $I_0(a)=I_{\rm tot}(a)h_0(a)/h_{\rm
  tot}(a)$ and $I(a,\tau)=I_{\rm tot}(a)h(a,\tau)/h_{\rm tot}(a)$ to
reconstruct these initial age distributions. These initial
distributions are plotted in Fig.~\ref{fig:several-delays}(a).  With
these initial conditions, we can solve Eq.~(\ref{eq:evol0}) and
Eq.~(\ref{eq:evol1}) with the method of characteristics to find the
full age and time dependence of the female populations

\begin{subequations}
\label{eq:sys4} 
  	\begin{align}
  		f_{0}(t,a) & =f_{0}(t-a,0)\exp\left[-\int_{0}^{a}\!\big(\mu_{\rm f}(a')
  		+\beta_{0}(a')\big)\dd a'\right]  \ \text{if }t>a
  		\label{eq:24a}\\
  		f_{0}(t,a) & =I_{0}(a-t)\exp\left[-\int_{a-t}^{a}\!\big(\mu_{\rm f}(a')+
  		\beta_{0}(a')\big)\dd a'\right]  \ \ \ \text{if }t\le a,
  		\label{eq:24b}
  	\end{align}
  \end{subequations}

  \begin{subequations}
	\label{eq:sys4b} 
	\begin{align}
		f(t,a,\tau) & =f(t-\tau,a-\tau,0)
\exp\left[-\int_{a-\tau}^{a}\!\big(\mu_{\rm f}(a')
		+\beta\left(a',a'-(a-\tau)\right)\big)\dd a'\right] \ \text{if }t>\tau
		\label{eq:25a}\\
		f(t,a,\tau) & =I(a-t,\tau-t)\exp\left[-\int_{a-t}^{a}\!\big(\mu_{\rm f}(a')
		+\beta\left(a',a'-(a-\tau)\right)\big)\dd a'\right] \ \ \ \ \ \  \text{if }t\le \tau.
		\label{eq:25b}
	\end{align}
\end{subequations}


%
Females are fertile only between sexual maturity and menopause. Thus,
we set $a_{\min}$ ($\sim 12$ years) and $a_{\max}$ ($\sim 50$ years),
so that $\beta_{0}(a)=\beta(a,\tau)=0$ for $a<a_{\min}$ or
$a>a_{\max}$. Recall that an imposed policy delayed-birth policy is
manifested by $\beta(a,\tau)=0$ for $\tau <
\delta$. Eq.~(\ref{eq:limCond0}) becomes
\begin{equation}
\label{eq:sys4-f00}
f_{0}(t,0)  =\eta \left(\int_{a_{\min}}^{a_{\max}}\!\!\beta_{0}(a)f_{0}(t,a)\dd a
+\int_{a_{\min}}^{a_{\max}}\!\!\!\dd a \int_{\delta}^{a}\!\!\dd\tau\, \beta(a,\tau)f(t,a,\tau)
\right).
\end{equation}
For $t\le \gamma \equiv \min\{a_{\min},\delta\}$, the $f_{0}(t,a)$ and
$f(t,a,\tau)$ terms in the integrands in Eq.~(\ref{eq:sys4-f00}) can
be solved by Eq.~(\ref{eq:24b}) and Eq.~(\ref{eq:25b}).  Thus, we can
express $f_0(t,0)$ in terms of $I_0(a), I(a,\tau), \beta_0(a),
\beta(a,\tau)$, and $\mu_{\rm f}(a)$. Using Eqs.~(\ref{eq:sys4}), we
can calculate $f_0(t,a)$ for any $a$ and $t\le \gamma$. Under the
imposed refractory period, Eq.~(\ref{eq:limCond1}) becomes
\begin{equation}
\label{eq:fsol}
 f(t,a,0)  =f_{0}(t,a)\beta_{0}(a)+ 
\int_{\delta}^{a}\!\!\beta(a,\tau)f(t,a,\tau)\dd\tau.
\end{equation}
If $t\le\gamma$, we can also use Eq.~(\ref{eq:25b}) for $f(t,a,\tau)$
in the integrand of Eq.~(\ref{eq:fsol}), and then use the solved
$f_0(t,a)$ to express $f(t,a,0)$ in terms of $I_0(a), I(a,\tau),
\beta_0(a), \beta(a,\tau)$, and $\mu_{\rm f}(a)$. Using
Eqs.~(\ref{eq:sys4b}), we can calculate $f(t,a,\tau)$ for any $a,\tau$
and $t\le \gamma$.  Finally, using $f_0(\gamma,a),f(\gamma,a,\tau)$ as
the initial conditions, we can solve $f_0(t,a),f(t,a,\tau)$ for $t\le
2\gamma$. Repeating this procedure, we can use
$I_0(a),I(a,\tau),\beta_0(a),\beta(a,\tau)$, and $\mu_{\rm f}(a)$ to
calculate $f_0(t,a)$ and $f(t,a,\tau)$ for any $t,a,\tau$.

Using the fundamental rates $\beta_{0}(a)$, $\mu_{\rm f}(a)$ and
$\beta(a,\tau)$ as those used in the previous subsection for the full
model (see Fig.~\ref{fig:example_birth-death-rates} and
Eq.~(\ref{eq:beta})), we use the above procedure to construct the
total female population $n(t)$ (see Eq.~\ref{NTOT}).  The evolution of
$n(t)$ over one century, under different interbirth delays, are
plotted in Fig.~\ref{fig:several-delays}(b). At long times, the total
population exhibits the asymptotic behavior predicted by the
eigenvalues shown in Fig.~\ref{fig:delays}(a). For $\delta \gtrsim
3.22$~years, the total population will decrease exponentially. Because
the total population growth rate is most sensitive to small values of
imposed delay $\delta$, even a delay of $\delta \sim 4$ years is
sufficient to dramatically reduce population over the next $100$ year,
compared to the $\delta=2$ case of no refractory period.

The formal results and analyses above can be generalized to include
time dependent parameters $\mu_{\rm f}(t,a)$, $\beta_{0}(t,a)$,
$\beta(t,a,\tau)$, and even $\delta(t)$ to reflect social and policy
changes. In this case, am imposed refractory period would be defined
by the time-dependent birth rate function $\beta(t,a,\tau) =
\beta_{0}(t,a)\mathds{1}(\tau > \delta(t))$ and the population
densities will need to be evaluated numerically.

\section{Results and Discussion}

Our basic structured population model can be modified and applied to
different scenarios and policies to make predictions about a number of
potentially relevant quantities. We focus on the population dynamics
in China under different control scenarios, paying particular
attention to age and sex distributions.

\subsection{Predictions and comparison to data from China}

\begin{figure}
	\centering{}\includegraphics[width=5.4in]{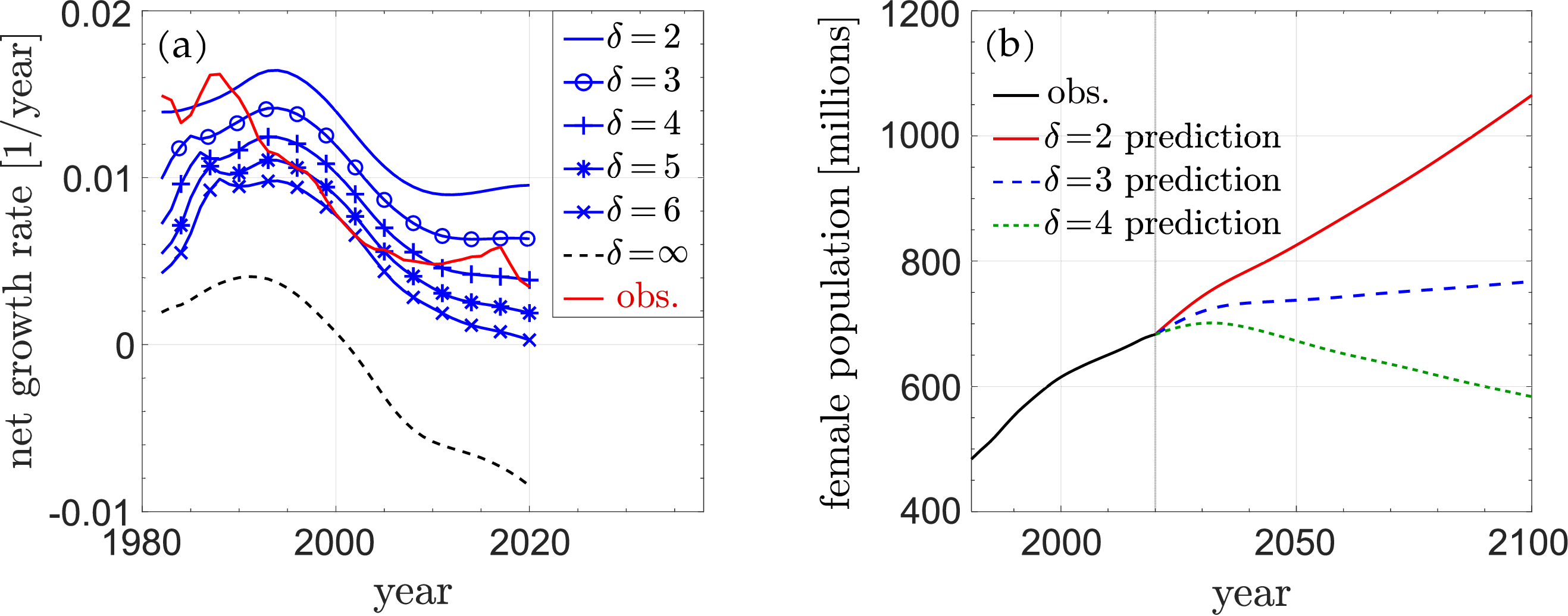}
	\caption{\label{fig:5} Predictions and comparison with real
          data. (a) Net growth rates in 1981-2020, observed (``obs'')
          and predicted growth rates (starting from 1981 conditions)
          under different policies $\delta$. (b) Predictions of the
          female population in 2021-2100 under different values of
          $\delta$.}
\end{figure}
First, we use parameters inferred from 1981 Chinese data in our model
to predict population growth for different values of $\delta$.  When
we compare predicted net growth rates with those derived from
1981-2020 data, shown in Fig.~\ref{fig:5}(a), we see that (i) during
1981-1990, \added{the observed growth rate is close to that when
  $\delta=2$}\footnote{\added{This means that there was effectively no
    interbirth period policy and that the the adjusted birth rate
    $\beta_0(a)$ was not changed much during the lax birth-control
    policies during this period \citep{li2020advocacy}.}}; \added{(ii) from
  1991 to 2010, the observed growth rate was close to that predicted
  from a model with $\delta=5$, possibly due to harsher policies like
  coerced abortion \citep{nie1999problem}; (iii) after 2011, the
  observed growth rate rose to a level close to that of a model with
  $\delta=3 \sim 4$, indicating a \textit{de facto} relaxation of the
  one-child policy. Indeed, after 2011, policies that
  \textit{encouraged} births were initiated.  Starting in 2011, a
  couple was allowed to have up to two children if both parents never
  had siblings.  Then, starting in 2014, a couple can have up to two
  children if at least one of the parents never had a
  sibling. Finally, starting in 2016, couples could bear up to two
  children regardless of their sibling status
  \citep{kane2021fertility}.  These policies might explain the
  flattening and subsequent increase in the net growth rate starting
  about 2011. However, the effects of these policies might be
  temporary. After the two-child policy in 2016, the net growth rate
  increased to the level consistent with a $\delta = 3$ model but soon
  returned to the level closer a $\delta = 4$ model. The actual growth
  rate is higher than in the $\delta=\infty$, strict one-child policy
  model. This indicates that many couples had, legally or illegally,
  more than one child.}

Starting in 2021, a new policy allows any couple to have up to three
children without penalty. We make different predictions for the effect
of this three-child policy. The optimistic prediction is that the
policy can fully stimulate childbearing to the point that the net
growth rate can reach the level predicted by a $\delta=2$ model, but
this could be realized only if behavior and cultural-economic changes
have not affected an intrinsic propensity for childbearing.  Another
possibility is that the policy has no significant effect so that the
net growth rate will only increase modestly and transiently before
effectively reducing back to that consistent with a $\delta=4$ model.
See Fig.~\ref{fig:5}(b) for different predictions of population over
the 2021-2100 time frame.

\subsection{Minimum childbearing age and population aging}

Besides mandating a refractory period between births, another method
of population control is to impose a minimum childbearing age.  The
minimum age $a_{\rm min}$ is thus set by policy rather than by
physiology.  For example, starting in 1985, in Yicheng county, Shanxi
province, a couple could bear two children, but the first child was
allowed only after the mother turned 24, and the second child was
allowed only after the mother turned 30 \citep{qin2017too}.\footnote{In
  China, the legal marriage age for females is 20 implying an
  existing soft constraint of $a_{\rm min}\approx 21$.}
\begin{figure}[H]
\centering{}\includegraphics[width=5.4in]{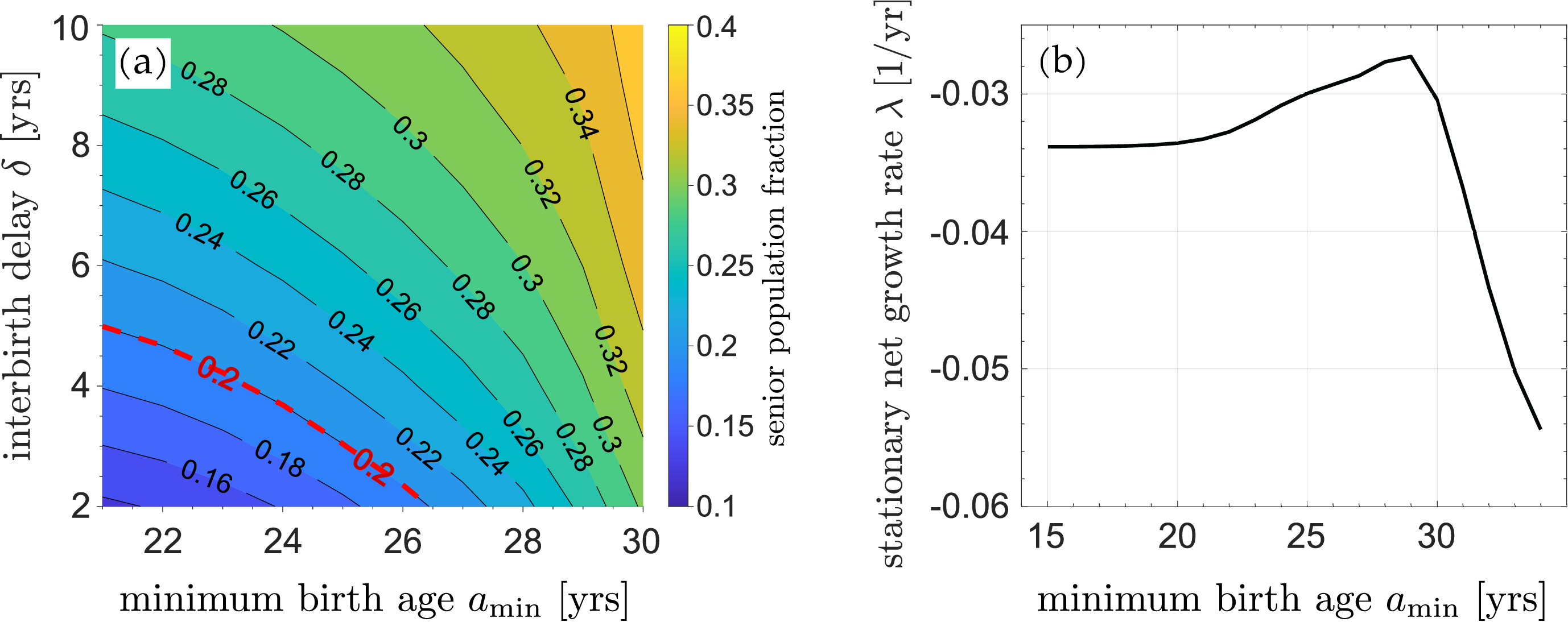}
\caption{\label{fig:6} The effect of adjusting the minimum
    childbearing age. (a) Increasing the interbirth delay $\delta$ and
    increasing the minimum childbearing age $a_{\min}$ have similar
    side effects of increasing the percentage of senior
    population. The red dashed line indicates a dangerous senior
    population percentage, $20\%$. (b) Under the strict one-child policy
    (namely, $\delta=\infty$), when we increase the minimum
    childbearing age $a_{\min}$, the stationary net growth rate
    $\lambda$ will first increase then decrease.}
\end{figure}

One side-effect of population control is a distribution shifted toward
older ages. In China, the percentage of seniors (65+) increased from
$5\%$ in 1981 to $13\%$ in 2020 \citep{man2021exploring}.  Policies
such as imposing interbirth delays $\delta$ and minimum childbearing
ages $a_{\min}$ can both affect the long term senior (65+) population.
Fig.~\ref{fig:6}(a) shows a contour plot of the percentage of seniors
as a function of imposed $\delta$ and $a_{\min}$. In order to maintain
the senior population under $20\%$ (the red dashed line), the overall
policy must not be too drastic.  Nonetheless, with two strategies, a
balanced combination can be used. For example, one can set
$a_{\min}=26,\delta=2$, or $a_{\min}=24,\delta=4$, or $a_{\min}=21,
\delta=5$ and still prevent the senior population from exceeding
$20\%$ far in the future.

Although increasing the minimum childbearing age $a_{\min}$ should
reduce the rate of childbirth, we observe a counter-intuitive scenario
in which the net growth rate is nonmonotonic in $a_{\min}$. Under the strict
one-child policy (i.e., $\delta=\infty$), increasing $a_{\min}$ will
first \textit{increase} the stationary net growth rate $\lambda$
before decreasing it, as shown in Fig.~\ref{fig:6}(b). Under a
perpetual strict one-child policy, the population in each successive
generation is roughly halved and the total number of future newborns
is roughly the current population.  Although decreasing $a_{\min}$ can
temporarily increase the total population, it accelerates the ``halving
process'' in the long run since the interval between successive
generations is shorter. When $a_{\min}$ is not large, almost every
woman can have one child anyway.  Further increasing $a_{\min}$ will
decrease $\lambda$ as more women start to be pushed past their
childbearing years without giving birth; thus, a maximum in the growth
rate $\lambda$ arises at $a_{\min}\approx 29$.

\subsection{Female population fraction and interbirth delay}

We used $\eta=0.48$, the fraction of female births 1981 China, to
generate the results presented in Section \ref{sec2}. Due to
subsequent sex-selective abortions biased towards males, the value of
$\eta$ dropped to $0.45$ in 2005 before gradually increasing
\citep{RATIO1992,goodkind2011child,RATIO2014}.
\begin{figure}[H]
	\centering{}\includegraphics[width=3.4in]{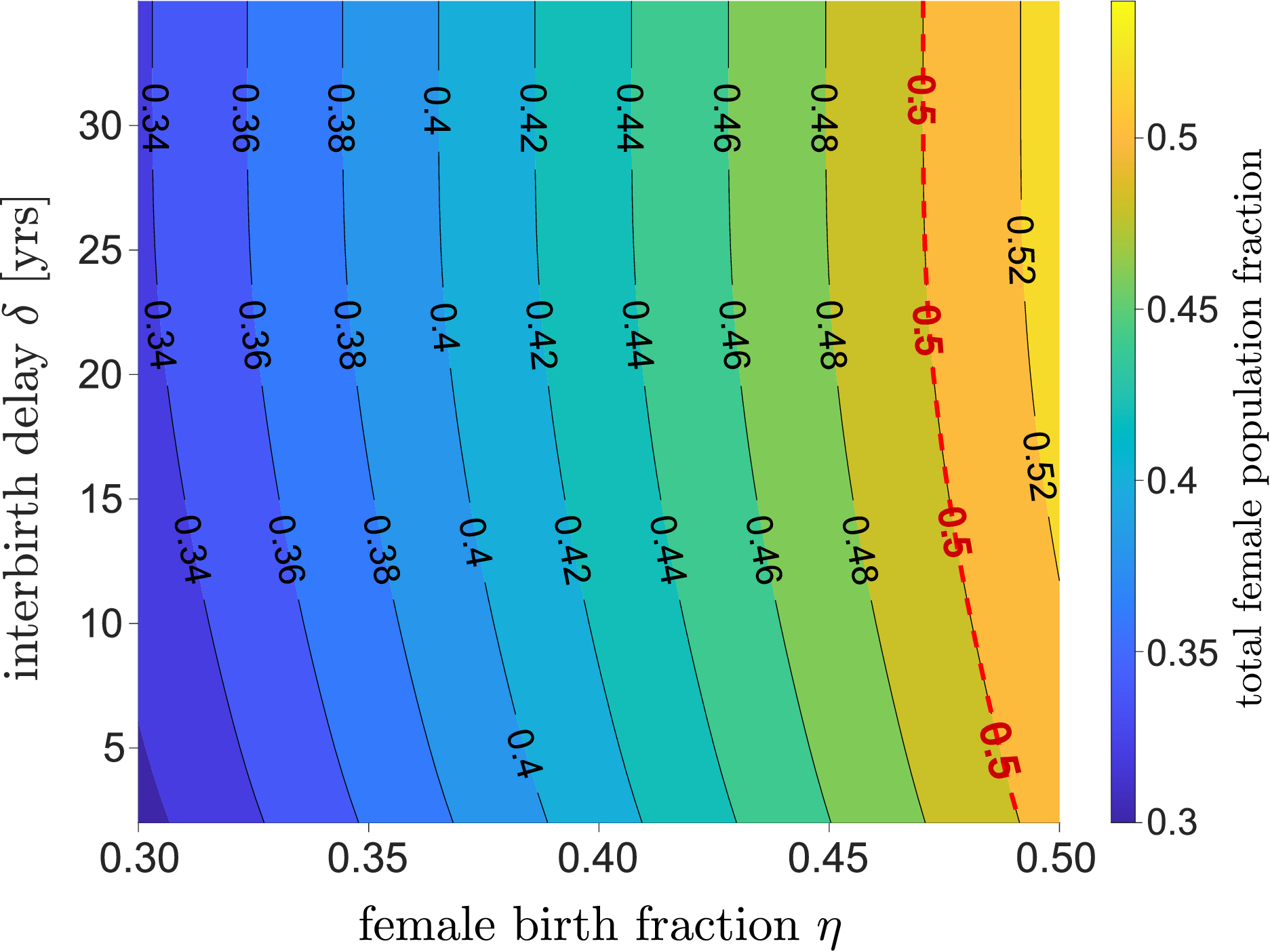}
	\caption{\label{fig:7} Dependence of the stationary female
          fraction on $\eta$ and $\delta$. Increasing the fraction of
          female births $\eta$ and the interbirth delay $\delta$ can
          both increase the stationary female population fraction. The
          red dashed line indicates conditions for 50\% females.}
\end{figure}
We can alter the value of $\eta$ and calculate the stationary female
fraction of total population, which is also affected by the interbirth
delay $\delta$. Fig.~\ref{fig:7} illustrates the stationary female
percentage as a function of $\eta$ and $\delta$. When $\delta=2$, the
stationary female percentage is approximately 1\% higher than the
female percentage $\eta$ at birth.  When we fix $\eta$ and increase
$\delta$, the stationary female fraction also increases. When
$\delta=\infty$, the stationary female population is approximately 3\%
higher than $\eta$.  For larger $\delta$, the stationary age
distribution is shifted to larger ages. Since the female death rate
$\mu_{\rm f}(a)$ is lower than the male death rate $\mu_{\rm m}(a)$ at
larger ages $a$, the female percentage increases with age (in 1981
China, newborns were 48\% female while 65+ seniors were 56\% female).


\subsection{Behavioral response to policies}
We have discussed the policy of applying a refractory period between
births and predicted its effects. We assume that the birth rate
returns to normal after the refractory period, meaning that people
obey this policy and do not respond with compensatory behaviors.  In
reality, people who want to have more children might mitigate the
effects of control policies by, for example, giving birth again
immediately after the end of the refractory period following the
previous birth. In addition to this ``catch up'' strategy to recover
from the ``missed opportunity,'' people might also prefer to have the
first child earlier, so that the refractory period finishes at an
younger age.

We propose a model that considers possible behavioral responses and
compare it with a no-response model. (1) For females of age $a$ who
have just finished their refractory period, the birth rate for the
following year (only) will be set to $\beta_{0}(a)c_1$ instead of
simply $\beta_{0}(a)$.  We can model the compensatory increase of
birth rate after the refractory period by
$c_1=1+0.1\times\min\{\delta-2,10\}>1$.  When the refractory period
$\delta$ is longer, people are more likely to more quickly make up for
the lost opportunity.  (2) For females of age $a$ who have not had
children, if $a+\delta\le 40$, the birth rate, instead of simply being
$\beta_{0}(a)$, will be set to $\beta_{0}(a)c_2$, where
$c_2=1+0.05\times\min\{\delta-2,10\}$. This means that females prefer
to have the first child earlier, if they know that they are young
enough to have another child after the refractory period (the age will
be $a+\delta$ at that future time).

\begin{figure}[H]
	\centering{}\includegraphics[width=5.4in]{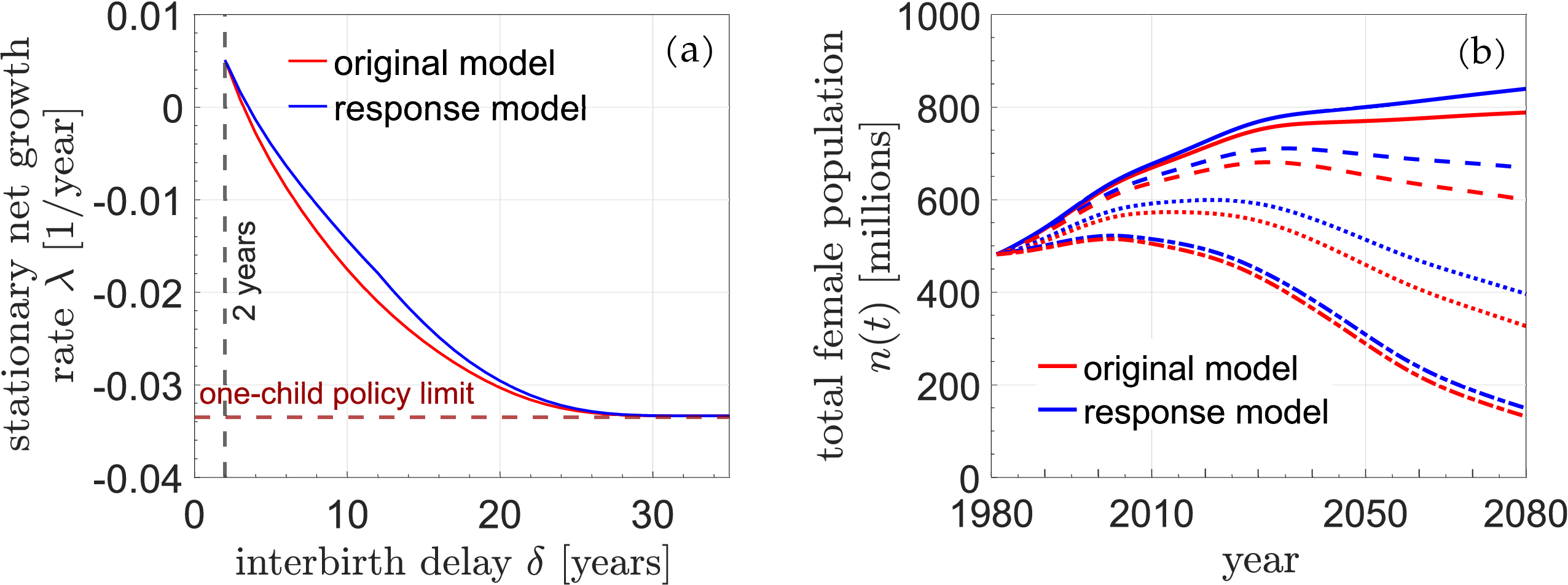}
	\caption{\label{fig:8} Comparison of the
            predictions of models with and without behavioral
            response. (a) Effect of an imposed interbirth delay
            $\delta$ on the asymptotic growth rate $\lambda$. The red
            curve depicts the no-response model, the same as the curve
            in Fig.~\ref{fig:delays}(a). The blue curve is the growth
            rate associated with the behavioral response model (where
            $\beta_{0}(a)c_1$ and $\beta_{0}(a)c_2$ are used as birth
            rates). For a moderate refractory periods ($\delta = 5\sim
            10$), the behavioral response is equivalent to making
            $\delta$ one year shorter. (b) Evolution of the total
            female population $n(t)$ for different imposed delays
            between births. The red and blue curves represent
            populations associated with the no-response (same as in
            Fig.~\ref{fig:several-delays}(b)) and behavioral response
            models, respectively. The solid, dashed, dotted,
            dash-dotted curves correspond to $\delta=3,4,7,15$ years,
            respectively. Behavioral responses have stronger effects
            on the total population when $\delta$ is not too large.}
\end{figure}

\added{Fig.~\ref{fig:8} compares predictions from the standard
  no-response model (red) to those from a behavioral response model
  (blue).  As expected, behavioral responses blunt the policy-induced
  decreases in the stationary growth rate (Fig.~\ref{fig:8}(a)) and
  the total population (Fig.~\ref{fig:8}(b)), resulting in
  higher-than-expected growth and populations. For an imposed
  $\delta$, a compensatory behavioral response model leads to a higher
  stationary growth rate. In other words, if the behavioral responses
  of this example are included, the imposed delay $\delta$ would have
  to be about 1-2 years longer than in the absence of behavioral
  response in order to achieve the same overall growth rate (for
  intermediate delays $\delta\approx 5-15$ years).  However, if the
  refractory period is set very long ($\delta\gtrsim 20$ years), our
  proposed behavioral responses are futile since females are
  irreversibly moved past their fertility window.}

\subsection{Comparison between China and Japan}
We have examined the effect of applying a refractory period policy in
China, which has implemented various birth-control policies over past
four decades. To better study this interbirth delay, we apply our
model to Japan, which does not have enforced polices on population
control. We use Japan's 2000 population data as a starting point
\citep{Census2000}. For Japan, we use its 2000 birth rate, which was
much lower than that of 1981 China.
	
\begin{figure}[H]
\centering{} 
\includegraphics[width=2.8in]{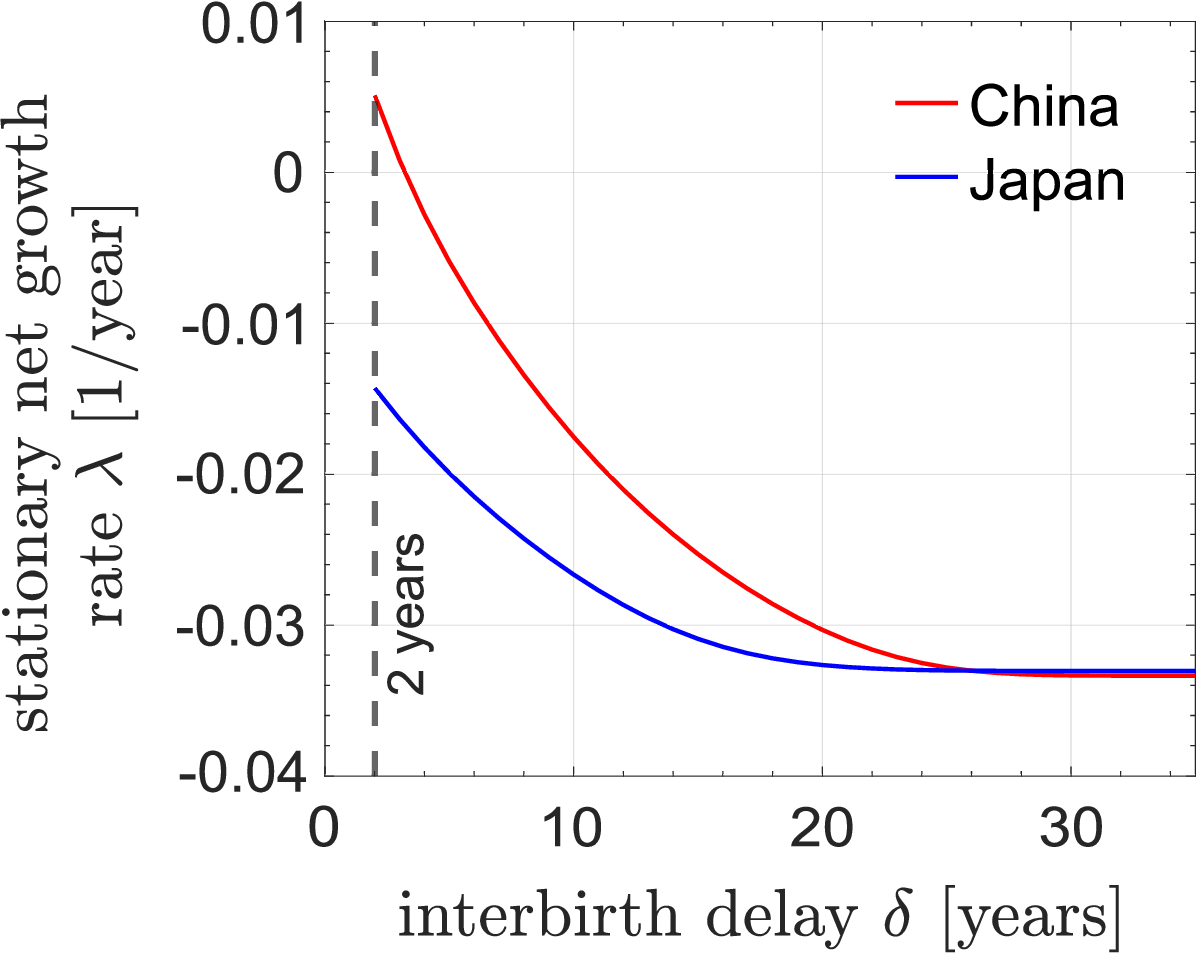}
\caption{\label{fig:9} Effect of an imposed interbirth delay $\delta$
  on the asymptotic growth rate $\lambda$, using birth rate data from
  China and Japan. The red curve is the asymptotic growth rate
  calculated from 1981 China data, the same as that shown in
  Fig.~\ref{fig:delays}(a). The blue curve is the asymptotic growth
  rate calculated from 2000 Japan data. Since Japan has a much lower
  birth rate, even without a refractory period policy ($\delta=2$),
  the stationary growth rate is negative. When $\delta$ is large
  enough, each female can have at most one child, and the asymptotic
  rate of population decay for China and Japan are nearly the
  same. However, under this large-$\delta$ sub-replacement limit, the
  long term growth rate for Japan can actually be slightly larger
  (less negative) than that of China.}
\end{figure}

Fig.~\ref{fig:9} compares the stationary growth rates between China
and Japan, imposing different refractory periods $\delta$.  Since
Japan has a much lower growth rate $\beta_0$, with the same $\delta$,
the stationary growth rate of Japan is lower than that of China. When
$\delta$ is sufficiently large, since each female can have at most one
child, the difference between China and Japan diminishes. In fact, the
limiting high-$\delta$ stationary growth rate of Japan is slightly
\textit{higher} than that of China.  Since the childbearing age is
older in Japan, the gap between two successive generations is
longer. As observed in Fig.~\ref{fig:6}(b), under large-$\delta$,
sub-replacement conditions, a moderately longer gap between
generations can increase the stationary (very long term) growth rate.
	
\section{Summary and Conclusions}

We have formulated a ``continuum'' of birth-control policies for
population management in which the strict one-child policy is a
limiting case. Our approach is based on explicitly incorporating a
refractory period between births. \added{In general, our age- and
  gestation-period structured model can also apply to organisms} in
which the gestation time is appreciable compared to an organism's
window of fertility.  For example, animals such as the Greater cane
rat (\emph{Thryonomys swinderianus}), the Pacarana (\emph{Dinomys
  branickii}) and the Steenbok (\emph{Raphicerus campestris}) have
gestation periods approximately $15\%$ of their fertility window
\citep{RATS}. Although a single gestation period is about $3\%$ of the
childbearing period in humans \citep{Tacutu2018}, socially imposed
refractory periods can be much longer (and infinite in a strict
one-child policy). Thus, our model provides a natural way to test how
imposed tunable interbirth refractory periods $\delta$ affect the
predicted total female population and its steady-state age
distribution. For long delays $\delta$, the model approaches the
strict one-child policy as a larger fraction of women are pushed past
menopause.

\added{In our mathematical analysis, we found a number of analytic or
  closed-form solutions to relevant demographic quantities such as the
  steady-state age distribution. We then considered an alternative
  scenario in which ``lax'' birth control policies that was being
  implemented in 1981 are kept, along with an additional policy of an
  imposed refractory period between births. Using 1981 as the starting
  point, we predicted population levels and compared them to the
  actual, realized populations. By applying a refractory period
  $\delta$ between births and using 1981 China birth rates, we
  provided a retrospective analysis and arrived at a number of
  quantitative conclusions.  Our analyses assumed that the birth rate
  $\beta_0$ and death rate $\mu$ did not change in the intervening
  years and that the population adhered to the birth-control policies
  without further behavioral responses.  We concluded that: (i) When
  $\delta\gtrsim 3.2$ years, the total population will not grow in the
  long run (Fig. \ref{fig:delays}(a)); (ii) When $\delta\ge 4$ years,
  the total population in China would have always been maintained
  under 1.45 billion (Fig.~\ref{fig:several-delays}(b)); (iii) When
  $\delta\ge 6$ years, the net growth rates during 1990-2010 (when a
  harsher one-child policy was applied) would be as low as what was
  realized (Fig.~\ref{fig:5}(a)); (iv) Without increasing the minimum
  childbearing age, when $\delta\le 5$ years, the stationary senior
  population would be maintained under $20\%$
  (Fig. \ref{fig:6}(a)).}

 \added{Such predictions assume the adjusted birth rate $\beta_0(a)$
   does not change over time. This assumption is definitely
   unrealistic, since many important socio-economic factors can affect
   birth rate distributions. The decrease of birth rate in China
   (illustrated in Fig. \ref{fig:several-delays}(a)) is not due solely
   to birth-control policies.  After 1980, female education increased,
   which had the statistically significant effect of decreasing the
   birth rate \citep{lan2016impact}. Additional evidence consistent
   with an extra-policy influence on birth rates in China is the
   increase in the average age of first childbirth (which one expects
   to be less affected by policies) from 24.3 years to 26.9 years from
   2006 to 2016 \citep{he2019china}. Moreover, we expect behavioral
   responses to policies that could mitigate their effectiveness.
   Since it is difficult to separate and quantify the effects of
   socio-economic factors and behavioral responses on birth rates, we
   did not explicitly incorporate these factors in our
   model. Nonetheless, we discussed how policies can be implemented
   through different modifications of age- and refractory
   period-dependent birth rate functions. For example, we considered a
   population-control policy whereby a minimum birth age $a_{\rm min}$
   is imposed. Here, we found the counterintuitive result that under a
   strict one-child policy, increasing $a_{\min}$ first
   \textit{increases} the stationary net growth rate, before
   decreasing it as $a_{\min}$ is further increased.}

 \added{Age-structured models can also be generalized to include
   additional subpopulations, such as those arising in cell division
   \citep{XIA_SIAP} and disease propagation \citep{RUAN2021}
   models. For example, in the birth control context, different
   generations and family structure can be enumerated in order to
   predict the effects of policies such as those implemented in 2011
   and 2014 that consider the sibling status of would-be parents,
   allowing those without siblings more latitude in
   childbirth. Additional concepts from sociology and response to
   socioeconomic and political influences can also potentially be
   integrated for a more complete framework of population dynamics and
   demography. The ideas and mathematical tools in this paper can be
   adapted to other fields. For example, an economist or a sociologist
   might study the cultural norms regarding child spacing and use our
   models to connect child spacing to growth rates.}



\section*{Data accessibility}
Data and relevant code for this research work are stored in GitHub:\\
\noindent \url{https://github.com/YueWangMathbio/ChildPolicy}\\
\noindent  and have been archived within the Zenodo repository:\\
\noindent \url{https://doi.org/10.5281/zenodo.6394805}.

\section*{Authors' contributions}
All authors collected and reviewed the literature, developed the model
and analysis, and wrote drafts of the manuscript. YW and RD analyzed
data and developed the computational methodology.  All authors
contributed to visualization and editing of the manuscript.  All
authors gave final approval for publication and agreed to be held
accountable for the work performed therein.

\section*{Competing interests}
We declare we have no competing interests.

\section*{Funding}
This work was supported by grants from the NIH through grant
R01HL146552 and the Army Research Office through grant
W911NF-18-1-0345.  The authors also thank the Collaboratory in
Institute for Quantitative and Computational Biosciences at UCLA for
support to RD.

\appendix



\part*{Mathematical Appendices}
\section{The equation of $f_{\rm tot}(t,a)$}
\label{ap:ftot}
We show by direct substitution that the total female population
density $f_{\rm tot}(t,a)$ satisfies the standard age-structured
McKendrick equation. Substitution of Eq.~\ref{eq:def_f_tot} into
Eq.~(\ref{eq:f_tot}) and expanding,

\begin{equation} 
	\label{eq:prf_tot}
	\begin{split}	 
\frac{\partial}{\partial t}f_{\rm tot}(t,a) 
+\frac{\partial}{\partial a}f_{\rm tot}(t,a) = &  
\frac{\partial}{\partial t}f_{0}(t,a)+\frac{\partial}{\partial a}f_{0}(t,a) \\
\: & +\int_{0}^{a}\!\frac{\partial}{\partial t}f(t,a,\tau)\dd \tau+
\int_{0}^{a}\!\frac{\partial}{\partial a}f(t,a,\tau)\dd \tau\\
\: & + f(t,a,a) +\int_{0}^{a}\!\frac{\partial}{\partial \tau}f(t,a,\tau)\dd \tau
-\int_{0}^{a}\!\frac{\partial}{\partial \tau}f(t,a,\tau)\dd \tau\\
= & -\left(\mu_{\rm f}(a)+\beta_{0}(a)\right)f_{0}(t,a)
-\int_0^a \big(\mu_{\rm f}(a)+\beta(a,\tau)\big)f(t,a,\tau) \dd \tau\\
\: & +f(t,a,a)-f(t,a,a)+f(t,a,0)\\
= & -\mu_{\rm f}(a)f_{\rm tot}(t,a)-\beta_{0}(a)f_{0}(t,a) 
-\int_0^a\!\beta(a,\tau)f(t,a,\tau) \dd \tau\\
\: & +\beta_{0}(a)f_{0}(t,a)+ \int_{0}^{a}\!\!\beta(a,\tau)f(t,a,\tau)\dd\tau\\
= & -\mu_{\rm f}(a)f_{\rm tot}(t,a).
	\end{split}
\end{equation}

\section{Relation between $\beta_{\rm eff}(a)$ and $\beta_{0}(a)$}
\label{ap:beb0}

In this Appendix, we prove Eq.~(\ref{eq:b0be}), the link between
$\beta_{\rm eff}(a)$ and $\beta_{0}(a)$. In the following, assume
$\tau\le \delta$. In Eq.~(\ref{eq:hat}), $\beta(a',a'-a+\tau)=0$ for
any $a-\tau < a'<a$. Thus, we have

\begin{equation}
\frac{h(a,\tau)}{h(a-\tau,0)}
=\exp\left[-\tau\lambda-\int_{a-\tau}^{a}\mu_{\rm f}(a')\dd a'\right],
\end{equation}
while from Eq.~(\ref{eq:h_lambda}), we have
\begin{equation}
\frac{h_{\rm tot}(a)}{h_{\rm tot}(a-\tau)}=
\exp\left[-\tau\lambda-\int_{a-\tau}^{a}\mu_{\rm f}(a')\dd a'\right].
\end{equation}
Thus,
\begin{equation}
\label{eq:hhhh}
\frac{h_{\rm tot}(a)}{h_{\rm tot}(a-\tau)}
=\frac{h(a,\tau)}{h(a-\tau,0)}.
\end{equation}
From Eq.~(\ref{eq:limCond1-1}) and Eq.~(\ref{eq:b_eff}), we have
\begin{equation}
\label{eq:bhh}
h(a-\tau,0)=\beta_{\text{eff}}(a-\tau)h_{\rm tot}(a-\tau).
\end{equation}
Upon combining Eq.~(\ref{eq:hhhh}) and Eq.~(\ref{eq:bhh}), we arrive at
\begin{equation}
\label{eq:bhh2}
h(a,\tau)=\beta_{\text{eff}}(a-\tau)h_{\rm tot}(a).
\end{equation}
Since Eq.~(\ref{eq:bhh2}) is valid for any $\tau\le \delta$, we have
\begin{equation}
\label{eq:ibhh}
\int_0^{\delta}\beta_{\text{eff}}(a-\tau)\dd \tau=
\frac{\int_0^{\delta}h(a,\tau)\dd \tau}{h_{\rm tot}(a)}.
\end{equation}
Eq.~(\ref{eq:b_eff}) can be transformed into
\begin{equation}
\beta_{\rm eff}(a)=\beta_{0}(a)\frac{h_{0}(a)+\int_{\delta}^{a}\!
h(a,\tau)\dd\tau}{h_{\rm tot}(a)}=\beta_{0}(a)\left[1-\frac{\int_{0}^{\delta}\!
h(a,\tau)\dd\tau}{h_{\rm tot}(a)}\right].
\label{eq:2b_eff}
\end{equation}
Combining Eq.~(\ref{eq:ibhh}) and Eq.~(\ref{eq:2b_eff}), we obtain
$\beta_{\rm eff}(a)=\beta_{0}(a)
\left[1-\int_0^{\delta}\beta_{\text{eff}}(a-\tau)\dd \tau\right]$ and thus
\begin{equation}
\beta_{0}(a)=\frac{\beta_{\text{eff}}(a)}{1-\int_0^\delta 
\beta_{\text{eff}}(a-\tau)\dd \tau}.
\label{eq:bef}
\end{equation}


\bibliographystyle{plainnat}
\bibliography{bibliography}

\end{document}